\documentclass[useAMS,usenatbib]{mn2e}
\usepackage[dvips]{color,graphicx}

\title[Many-body interactions in dynamical friction]{Corrective effect of many-body interactions in dynamical friction}
\author[S. Inoue]{Shigeki
Inoue$^{1,2}$\thanks{E-mail:inoue@astr.tohoku.ac.jp}
\\
$^{1}$Astronomical Institute, Tohoku University, Sendai 980-8578, Japan\\
$^{2}$Mullard Space Science Laboratory, Holmbury St. Mary, Dorking, Surrey, RH5 6NT}

\begin{document}

\date{2010 September 27}

\pagerange{\pageref{firstpage}--\pageref{lastpage}} \pubyear{2002}

\maketitle

\label{firstpage}

\begin{abstract}
Dynamical friction is a fundamental and important phenomenon in astrophysics. The Chandrasekhar formula is a well-known analytical estimation of the effect. However, current astrophysicists have realized that the formula is not correct in some cases because of several approximations dared in the formulation and/or complex non-linearities in the real universe. For example, it has been indicated that the dynamical friction doesn't work in cored density profiles (constant density in the central region) despite that the Chandrasekhar formula predicts drag force even in the constant densities. In the former half of this paper, I discuss by N-body simulations that many-body interactions are also important in actual dynamical friction though derivation of the Chandrasekhar formula is based on the assumption of two-body interaction. In the simulation, the many-body interactions are caused by a very small number of field particles co-rotating with a perturber. However, the contribution from the many-body interactions accounts for a non-negligible fraction of the actual dynamical friction. In the latter half, I discuss why the cored profiles suppress the dynamical friction. One possible explanation is that corrective effect of the many-body interactions drive orbital motion of the perturber. The cessation of dynamical friction by this corrective effect would be feasible even in shallow cusp density profiles although the shallow cusp may evolve into a constant density.

\end{abstract}


\section{Introduction}

Dynamical friction (hereafter, DF) is an important physical process for a wide variety of dynamics, e.g. stellar dynamics, galaxy and planetary formation etc. That's why it is in great demand to seek deeper understanding of the nature of DF. As a historical milestone, \citet{c:43} has succeeded in formulation of the effect of DF. However, several approximations were unavoidable in the derivation of the formula. The Chandrasekhar formula was derived under the assumptions of uniform density and isotropic velocity field. Moreover, interaction between a perturber and field particles was supposed to be integration of countless two-body interactions and orbital periodicity of all particles was not taken into account. By efforts of many numerical studies, current astrophysicists have realized that the Chandrasekhar formula is not perfect and loses accuracy of the estimation in some cases \citep[e.g.][]{tw:84,w:89,jb:00,ffm:06}. 

In this paper, I concentrate on the suppression of DF on a globular cluster (hereafter, GC) in a cored halo of dwarf galaxies \citep{hg:98,gmr:06,srh:06,rgm:06,ad:09,cwb:09,i:09}. However, the discussion in this paper can be generalized to other astronomical bodies by scaling physical units. The drag force by DF is negligibly weak for GCs in large systems, like our Galaxy. In contrast, in small systems like dwarf galaxies, the drag force is far too strong \citep[see][chap. 8]{bt:08}. Thus, the GCs in dwarfs are presumed to lose their orbital energy and fall into the galactic centre by strong friction force in time-scale of the order of $\sim 1$ Gyr \citep{t:76,hg:98,olr:00,v:00,v:01,gmr:06,b:09}. Nevertheless, these GCs still do exist in many dwarf galaxies even now and are as old as the age of universe \citep[e.g.][]{dhg:96,mlf:98,bcz:98,bcc:99,sbf:03,mg:03,lmf:04,g:07}. This problem is referred to as `\textit{the dynamical friction problem}'.

However, a solution of the problem has been proposed. If dwarf galaxies have a cored dark matter halo which has a constant density region in its centre, the DF against the halo is weakened considerably, and GCs should be able to survive beyond the age of the universe. By an analytical approach using the Chandrasekhar formula, \citet{hg:98} discovered that a King-model halo can weaken the DF in the core region. Recently, by N-body simulations, \citet{gmr:06}, \citet{rgm:06} and \citet{i:09} confirmed complete cessation of DF in a core region of haloes. Thus, although the Chandrasekhar formula indicates gradual shrinkage of the GC orbit, the N-body simulations demonstrate complete defunctionalization of DF. This discrepancy between analytic and numerical studies (the Chandrasekhar formula and N-body simulations) still remains. Besides, the cessation of DF is a topic not restricted to dwarf galaxies and GCs. \citet{gm:08} has discussed the same core-stalling of DF on binary black hole against a cored stellar cluster and suggested that fluctuation of the halo potential weakens the DF in the core.

My goals in this paper are to clarify hidden physics in the nature of DF and the mechanism of the DF suppression in cored density profiles by N-body simulations. Constant density distribution is called, in other words, `harmonic oscillator potential', which means that all particles in the potential have the same orbital period. In this sense, harmonic potentials are expected to induce complex effects in non-linear manner and be beyond the applicable domain of the Chandrasekhar formula. It has been known for a long time that DF shows curious behaviors in constant densities. \citet{tw:84} analytically argued that a constant density core causes energy feedback on a perturber by dynamical resonance in positive or negative sense depending on the density distribution. In the same context as \citet{tw:84}, \citet{w:89} indicated that density response of the density field to a perturber orbiting in the core region can exert torque on the perturber and drive the orbital motion. Moreover, \citet{bpp:08} also analytically argued that in the harmonic potentials the perturber acts as a catalyst for the evolution of the energy distribution of the field particles by the orbital resonance.

By the way, the cored dark haloes are inconsistent with results of cosmological N-body simulations \citep[e.g.][]{nfw:97,swv:08}. However, among small scale galaxies, many observations of low surface brightness (LSB) galaxies have confirmed the cored dark haloes \citep[e.g.][and refernces therein]{b:95,sb:00,bmb:01,lc:09,d:10}, and nearby dwarf galaxies also favor the cored haloes \citep{gww:07}. In addition, numerical simulations including baryonic physics also predict core creation by bursty supernovae in the centre of dwarf haloes in cosmological context \citep{mwc:08,g:10} and isolated models \citep{pgb:10}. 

This paper is organized as follows: I will explain my simulation method and present my simulation results in \S 2. In \S 3, I will show analyses of the data generated by my simulations. In \S 4, I will summarize my results and present discussion.

\section{The simulations}
My numerical simulation is a typical pure N-body simulation (no gas component). I use Barnes-Hut modified tree-code \citep{bh:86,b:90}, setting an open angle of $\theta = 0.5$. A special-purpose calculator for collisionless N-body simulations, GRAPE-7, is used with the tree algorithm \citep{m:91}. The leapfrog time-integrator is adopted with the shared-timestep method. The total number of timesteps is 11841 for the whole of a simulated period which corresponds to 10 Gyrs in the real time-scale.

\subsection{The settings}
In this paper, I use Burkert profile \citep{b:95} as a cored halo model of dwarf galaxies, which is motivated from observations of LSB galaxies \citep{sb:00,lc:09} and numerical simulations of formation of dwarf galaxies \citep{mcw:06},
\begin{equation}
  \rho(r)=\frac{\rho_0r_0^3}{(r+r_0)(r^2+r_0^2)}.
  \quad\quad ;\;r<r_{vir}
  \label{Burkert}
\end{equation}
$\rho_0$ is the effective core density, $r_0$ is the core radius which defines the constant density region. In my simulation, I set $\rho_0=0.1M_\odot/pc^3$, $r_0=1.0$ kpc, and the virial mass of the halo $M_{vir}\equiv M(r_{vir})=2.0\times10^9M_\odot$, the virial radius $r_{vir}=9.88$ kpc. Eq.\ref{Burkert} estimates the mass enclosed inside 300 pc to be $M(300pc)=8.79\times10^6M_\odot$. This value is consistent with the result of \citet{sbk:08} which has suggested that all dwarfs have the same mass scale, $M(300pc)\sim10^7M_\odot$. I add an exponentially decaying envelope to prevent instability at the outer region caused by the artificial cut-off radius, $r_{vir}$ \citep{sw:99},
\begin{equation}
  \rho(r)=\rho(r_{vir})\biggl(\frac{r}{r_{vir}}\biggr)^a\exp{\frac{r_{vir}-r}{r_{decay}}},
  \quad\quad ;\;r>r_{vir}
  \label{envelope}
\end{equation}
where
\begin{eqnarray*}
 a&=&\frac{r_{vir}}{r_{decay}}+r\frac{d}{dr}\ln\rho\big\arrowvert_{r_{vir}}\\
&=&\frac{r_{vir}}{r_{decay}}-\frac{c}{1+c}-\frac{2c^2}{c^2+1},
 \label{a}
\end{eqnarray*}
and $c\equiv r_{vir}/r_0$. I set $r_{decay}\equiv0.1r_{vir}$. With this envelope, total halo mass enclosed in $r_{cutoff}=3.16\times10^4$ pc becomes $M_{total}=2.44\times10^9M_\odot$. 

Velocity dispersion is given by the solution of Jeans equations, 
\begin{equation}
 \sigma_r^2(r)=\frac{1}{r^{2\beta}\rho(r)}\int_r^{\infty}dr'r'^{2\beta}\rho(r')\frac{d\Phi}{dr'},
 \label{jeans}
\end{equation}
where $\beta$ is the anisotropy parameter. In this paper, I assume the isotropic velocity state in the halo, setting $\beta=0$ ($\sigma_r=\sigma_\theta=\sigma_\phi$). The velocity distribution is determined by local Maxwellian approximation,
\begin{equation}
 F(v)=4\pi\Bigl(\frac{1}{2\pi\sigma^2}\Bigr)^{3/2}v^2\exp{\Bigl(-\frac{v^2}{2\sigma^2}\Bigr)},
 \label{maxwellian}
\end{equation}
where $F(v)$ is a probability distribution function of velocities \citep{h:93}. By the assumptions of isotropic and Maxwellian velocity distribution, I satisfy the suppositions underlying the derivation of the Chandrasekhar formula, although \citet{lfd:05} has reported slight instability in the Burkert profile realized by the local Maxwellian approximation. 

I conduct N-body simulations by $10^7$ particles to resolve the halo. All halo particles have the same mass, $m=2.44\times10^2M_\odot$, with a softening length of $\epsilon=3.0$ pc. Before operating actual simulations, the halo model needs to be relaxed to some extent. I run a simulation without GCs for 2 Gyrs. After the relaxation, I employ the relaxed halo as the initial state ($t=0$) of the halo.

In my simulation, the GC is represented by a point mass with $m_{gc}=2.0\times10^5M_\odot$. The softening length of the GC particle is set to $\epsilon_{gc}=10$ pc. Initial orbital radius of the GC is set to 750 pc. The initial orbit is circular, and I employ a single GC in my simulation. Multi-GC cases are also interesting, but have been investigated by \citet{gmr:06} and \citet{i:09}, which didn't indicate any large differences from the single GC cases. If the GC is resolved by many particles, some additional effects may play important roles: mass-loss, dynamical heating, etc \citep[e.g.][]{ffm:06,mcd:06,ef:07,cm:08}. Besides, stellar components in a dwarf galaxy can enhance the DF of GCs \citep{srh:06,b:09}. However, in this paper, I concentrate on investigations of fundamental physics of DF, and do not consider such effects. 


\subsection{The results}
\begin{figure}
 \includegraphics[width=84mm]{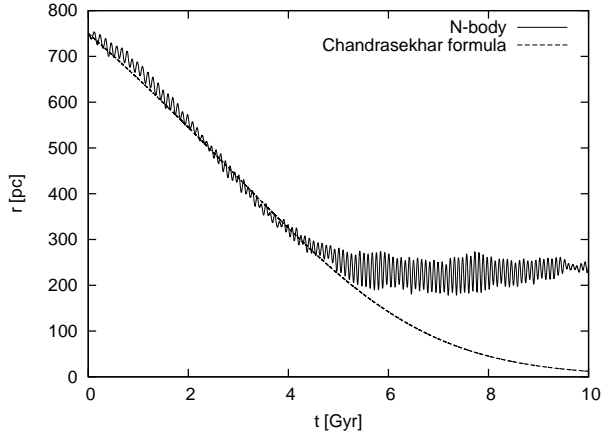}
 \caption{The time-evolutions of orbital radius of the GC. The Chandrasekhar formula is calculated with $\ln\Lambda=3.72$.}
 \label{fig1}
\end{figure}

The time-evolution of orbital radius of the GC is plotted in Fig.\ref{fig1}. As a comparison, I also plot an analytic calculation by the Chandrasekhar formula,
\begin{equation}
\frac{dv_{c}}{dt}=-\frac{4\pi\ln\Lambda G^2\rho m_{gc}}{v_{c}^2}\Bigl[\rm{erf}(X)-\frac{2X}{\sqrt{\pi}}e^{-X^2}\Bigr],
 \label{chandra}
\end{equation}
\begin{equation}
\frac{dr}{dt}=\frac{r}{v_{c}}\frac{dv_{c}}{dt},
 \label{drdt}
\end{equation}
where $X\equiv v_{c}/(\sqrt{2}\sigma)$. $v_{c}$ is circular velocity of the GC as a function of radius. $G$ is the gravitational constant. For the Coulomb logarithm, I set $\ln\Lambda=3.72$, which is determined by the best-fitting to the result of simulation for $t<4$ Gyr. If I take the original definition, $\ln\Lambda\equiv\ln(b_{max}/b_{min})=\ln(r_0/\epsilon)=5.81$, where $b_{max,min}$ are the maximum and the minimum impact parameter. In the derivation of Eq.\ref{drdt}, the assumption of circular orbit is used \citep{hg:98,i:09,cwb:09}. As seen from Fig.\ref{fig1}, in the N-body simulation, the orbital shrinkage stops after the GC entered into the constant density region ($r\hspace{0.3em}\raisebox{0.4ex}{$<$}\hspace{-0.75em}\raisebox{-.7ex}{$\sim$}\hspace{0.3em}300pc$). On the other hand, the Chandrasekhar formula fails to reproduce the N-body result in the core region, in spite of the excellent fit outside the core. In this paper, I define $t\hspace{0.3em}\raisebox{0.4ex}{$<$}\hspace{-0.75em}\raisebox{-.7ex}{$\sim$}\hspace{0.3em}5$ Gyr as `\textit{exerting DF phase}', and $t\hspace{0.3em}\raisebox{0.4ex}{$>$}\hspace{-0.75em}\raisebox{-.7ex}{$\sim$}\hspace{0.3em}5$ Gyr as `\textit{suppressed DF phase}'.

\begin{figure}
 \includegraphics[width=84mm]{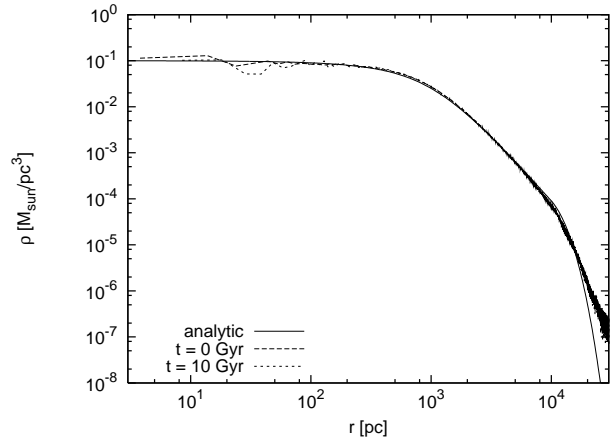}
 \caption{Density profiles of field particles of analytic model (Eq.\ref{Burkert} and \ref{envelope}) and in the N-body simulation for t=0 and 10 Gyr.}
 \label{fig2}
\end{figure}
The density profiles of the halo are plotted in Fig.\ref{fig2}. Even in the end-state of the simulation, the density profile is stable. The energy conservation rate of the system, $1-E_{end}/E_{ini}$, is $1.92\times10^{-4}$.

\section{The analysis}
\subsection{The exerting DF phase}
In the N-body simulation, for $t<5$ Gyr, DF is clearly exerted on the GC, leading to orbital shrinkage. The Chandrasekhar formula with $\ln\Lambda=3.72$ succeeds in estimation of the effect in this phase.

Firstly, I calculate dynamical energies of all particles. The specific energy of the $i$th particle is
\begin{equation}
  E_i=\frac{v_i^2}{2}-\frac{Gm_{gc}}{\sqrt{|\vec{x_{gc}}-\vec{x_i}|^2+\epsilon_{gc}^2}}-\sum^n_{j=1\atop{j\not=i}}\frac{Gm}{\sqrt{|\vec{x_j}-\vec{x_i}|^2+\epsilon^2}},
  \label{energy}
\end{equation}
where $\vec{x}_{i,j,gc}$ mean position vectors of the $i,j$th field particles and the GC, respectively.
\begin{figure}
 \includegraphics[width=84mm]{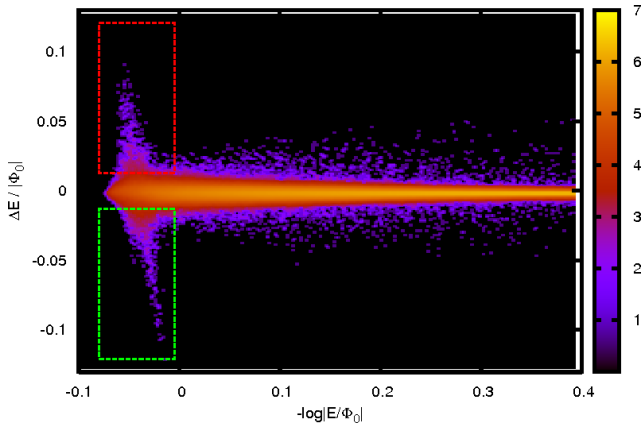}
 \caption{Contour plot of distribution of field particles in a $E-\Delta E$ diagram. The ordinate indicates an amount of increase of dynamical energy of each particle between 2 and 3 Gyr: $\Delta E\equiv E(3Gyr)-E(2Gyr)$. The abscissa indicates energy before the change: $E(2Gyr)$. The values are normalized by $|\Phi_0|$ which is absolute value of the potential energy at $r=r_0$ analytically estimated. Colour-bar labels mass in a bin, $\log(M/M_\odot)$. Black colour means no mass. In this diagram, I separate the particles into three groups; positive-horn (the red square), negative-horn (the green square), the others.}
 \label{fig3}
\end{figure}
Next, I calculate increase of $E_i$ at intervals of 1 Gyrs. In Fig.\ref{fig3}, I show a distribution of all field particles in $E$-$\Delta E$ diagram. $\Delta E$ corresponds to the increase of $E_i$ between 2 and 3 Gyr. In this figure, I can find a notable feature. At a certain energy, $-\log|E/\Phi_0|\simeq-0.05$, there are some particles of which energies changed largely. This feature looks like two \textit{horns}. Thus, I define $-\log|E/\Phi_0|<-5.23\times10^{-3}$ and $\Delta E/|\Phi_0|>1.27\times10^{-2}$ (inside the red square) as \textit{positive-horn} (P-horn), $-\log|E/\Phi_0|<-5.23\times10^{-3}$ and $\Delta E/|\Phi_0|<-1.27\times10^{-2}$ (inside the green square) as \textit{negative-horn} (N-horn). The numbers of P- and N-horn particles in this plot are 1,513 ($3.69\times10^5M_\odot$ in total) and 2,028 ($4.95\times10^5M_\odot$ in total), respectively. 

\begin{figure}
  \includegraphics[width=90mm]{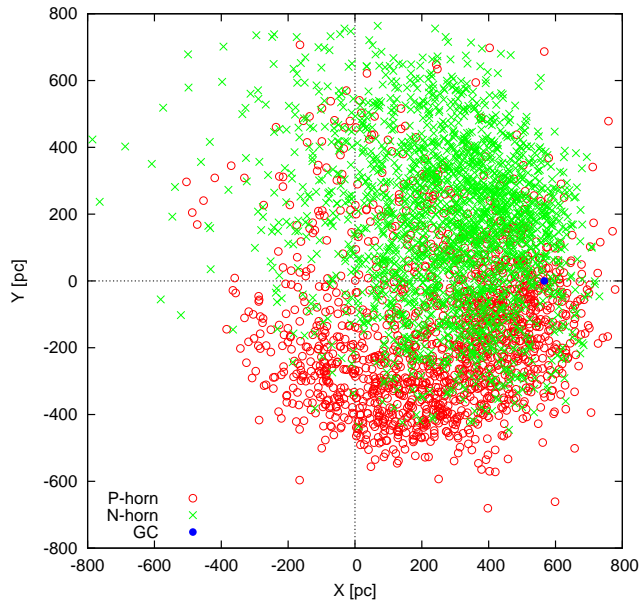}
 \caption{Positions of the P-, N-horn particles and the GC at $t=2$ Gyr. I set X-axis to the direction of the GC from the galactic centre. Z-axis coincides with the GC angular momentum vector. Y-axis lies in the GC orbital plane. The GC is rotating counter-clockwise in the X-Y plane.}
 \label{fig4}
\end{figure}
In Fig.\ref{fig4}, I show positions of the P- and N-horn particles at t=2 Gyr. The particles are projected on the GC orbital plane. By calculating dot products of angular momentum vectors of the GC and the horn particles, I confirm that most of the particles plotted have positive dot products at t=2 Gyr: prograde rotations with the GC (1,194 of 1,513 P-horn and 1,734 of 2,028 N-horn particles have prograde rotations). Thus, I find that most of the P-horn particles are orbiting behind the GC, whereas, the N-horn particles are orbiting ahead of the GC. Moreover, since most of these particles have short distances from the GC and the prograde rotations, I can ascribe the horn-feature in Fig.\ref{fig3} to the similarity between orbits of the horn particles and the GC. These particles strongly interact with the GC. Conversely, the GC ought to be affected by these particles. 

\begin{figure}
  \includegraphics[width=90mm]{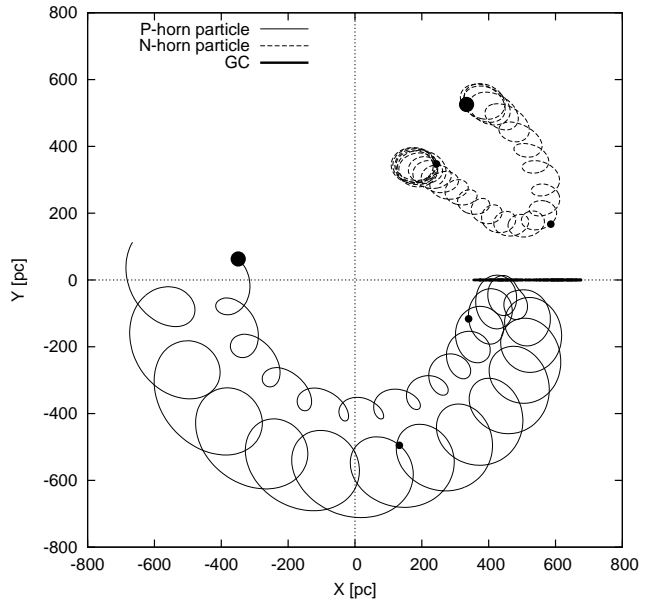}
  \caption{Examples of typical orbits of the P- and N-horn particle in rotating coordinate with the GC. Although the horn particles are defined by energy increase in t=2-3 Gyr, the orbits plotted here are for t=1-4 Gyr. The large (small) dots indicate positions of the particles at t=1 Gyr (t=2 and t=3 Gyr).}
  \label{RP_T2}
\end{figure}
I examine orbital motions of the horn particles. In Fig.\ref{RP_T2}, I illustrate typical orbits of the P- and N-horn particles in the coordinate rotating with the GC. The P-horn particle in this figure approaches the GC from behind. The orbit expands when catching-up, and finally retreats from the GC after the orbital expansion. This orbital expansion corresponds to the energy increase seen in Fig.\ref{fig3}. On the other hand, the orbit of the N-horn particle is the reverse, comes close to the GC (in inertia coordinate, the GC is catching up to the N-horn particle), shrinks the orbital radius, and runs away from the GC. As a result, the energy of the N-horn particle decreases. The mechanism of these orbital behaviors is simple. Under density gradients, generally, inner orbits have shorter orbital periods, on the other hand, outer ones have longer periods. The P-horn particle is pulled forward by gravity of the GC, and gains energy. However, this energy increase causes the orbital expansion and makes angular velocity slow. In the case of the N-horn particle, the particle is pulled backward and decrease the energy. This energy decrease results in the orbital shrinkage and faster angular velocity. This phenomenon, that the particle slows down in azimuth when pulled forward and speeds up when held back, has come to be called `the donkey effect' \citep{lk:72,bt:08}. The orbits in Fig.\ref{RP_T2} imply that higher energy particles lose orbital energy, on the other hand, lower energy particles gain energy. This behavior can be confirmed in Fig.\ref{fig3}, which indicates that the two horns lean slightly from top-left to bottom-right.

\citet{fif:09} has also investigated this kind of orbital motion (Trojan horseshoe orbits) in N-body systems, which look like the orbits in Fig.\ref{RP_T2} \citep[see][Fig.7]{fif:09}. It is worthy of special mention that these orbits exhibit quite different behavior from two-body interaction that is assumed in the derivation of the Chandrasekhar formula. This orbital behavior of the horn particles is presumed to be caused by many-body interaction, rather than two-body interaction (see, Appendix A). 

\begin{figure}
  \includegraphics[width=84mm]{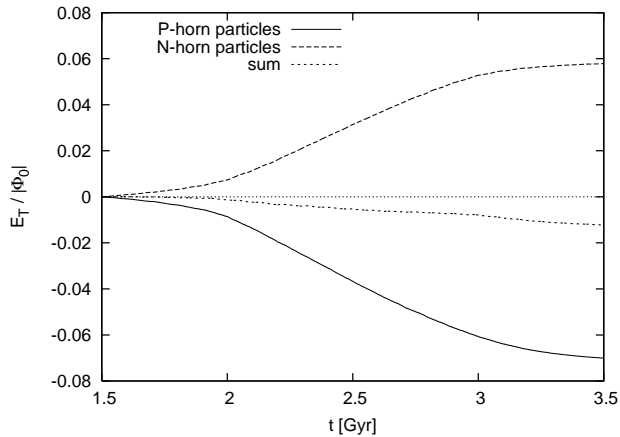}
  \caption{Cumulative energy transfers to the GC from the P- and N-horn particles in the exerting DF phase. The short-dashed line corresponds to the sum of both. Increase in $E_T$ means energy injection into the GC, decrease means energy absorption from the GC.}
  \label{fig5}
\end{figure}
To probe the energy transfer from the horn particles to the GC, I rerun the same simulation while calculating mechanical work on the GC,
\begin{equation}
  E_T(t)=\int^t_{t_{ini}}\vec{v_{gc}}(t')\cdot\vec{a}_{spc}(t')dt'.
  \label{dE}
\end{equation}
This integration means cumulation of the integrand on each timestep. $\vec{a}_{spc}$ represents net acceleration exerted on the GC by a specific particle group (e.g. the P-, N-horn or the other particles).

I compute energy transfers on the GC in the same simulation and show the result in Fig.\ref{fig5}. Though the horn particles were defined by energy change during 2-3 Gyr, I draw the energy transfers during  t=1.5-3.5 Gyr in Fig.\ref{fig5}; setting $t_{ini}=1.5$ Gyr in Eq.\ref{dE}. In agreement with a naive expectation, I find that the P-horn particles absorb energy from the GC and the N-horn particles inject energy into the GC. It is notable that the behaviors of the injection and absorption are monotonic. The transfers are sharp in 2-3 Gyr, slow in $t\hspace{0.3em}\raisebox{0.4ex}{$<$}\hspace{-0.75em}\raisebox{-.7ex}{$\sim$}\hspace{0.3em}2$ Gyr and $t\hspace{0.3em}\raisebox{0.4ex}{$>$}\hspace{-0.75em}\raisebox{-.7ex}{$\sim$}\hspace{0.3em}3$ Gyr. This implies that the energy transfers are transient.

\begin{figure}
  \includegraphics[width=84mm]{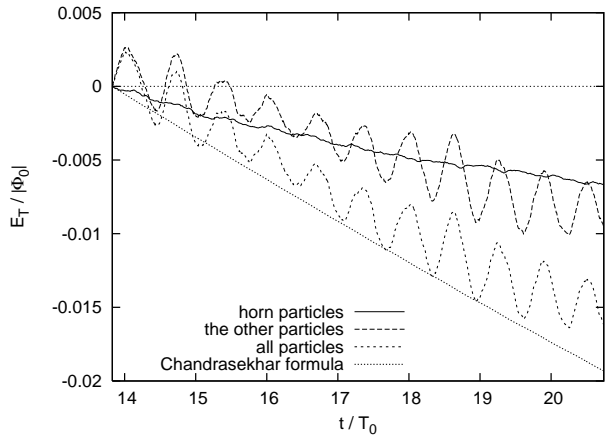}
  \caption{Comparison of cumulative energy transfers to the GC. The solid line labelled as `horn particles' is the same as `sum' in Fig.\ref{fig5} but setting $t_{ini}=2$ Gyr in Eq.\ref{dE}.}
  \label{fig6}
\end{figure}
The energy transfers in Fig.\ref{fig5} don't completely balance; however, the energy absorption by the P-horn particles slightly surpasses the injection by the N-horn. I compare the energy transfer on the GC by both of the P- and N-horn particles with that by the other particles in Fig.\ref{fig6}. The figure indicates that the other particles also steal energy from the GC. This energy loss of the GC seems to be caused by DF against the other field particles except the horn particles. Interestingly, the energy loss caused by the horn particles is comparable to that by the other particles, accounting for a half of actual DF by all particles. The dotted line labelled as `Chandrasekhar formula' is calculated by substituting Eq.\ref{chandra} for $\vec{a}_{spc}$ in Eq.\ref{dE} using the data of GC orbit in the simulation. The DF from all particles is well fitted by the Chandrasekhar formula with $\ln\Lambda=3.72$ This result implies that actual DF can be separated into two modes. The first mode is the well-known conventional DF based on the concept of \citet{c:43}. This is a DF that is attributed to random encounters with a myriad of field particles in an isotropic velocity distribution. On the other hand, unlike the first mode, the cause of the second mode is a small number of particles which are in many-body interaction; the horn particles. The horn particles have only 0.0354 per cent of the total mass of the system.

However, keep in mind that the criterion for defining the horn particles is somewhat arbitrary. Moreover, the result in Fig.\ref{fig6} doesn't necessarily mean that DF caused by `\textit{true}' many-body interacting particles accounts for a half of the actual DF. Even in two-body interaction, if a field particle has a similar velocity vector to $\vec{v_{gc}}$ and a small impact parameter, the particle causes strong interaction and large energy transfer. There would be no clear-cut way to extract only the true many-body interacting particles. However, it is sure that the horn particles defined in Fig.\ref{fig3} include many many-body interacting particles. 

\subsection{The suppressed DF phase}
For $t>5$ Gyr, the DF doesn't seem to act on the GC to shrink the orbit. This phenomenon itself has already been known. \citet{hg:98} found that this core-stalling of DF can be explained by the Chandrasekhar formula. However, their estimation was based on the assumption of completely constant density in the core. Without this assumption, the Chandrasekhar formula predicts slow but certain orbital shrinkage \citep{gmr:06,rgm:06,i:09}, as seen in Fig.\ref{fig1}.

On the other hand, the N-body simulations indicate the complete suppression of the DF in the cored region. The Chandrasekhar formula is no longer viable in this phase. \citet{rgm:06} has suggested that the mechanism of the DF suppression is anisotropic velocity distribution induced by the GC. They attributed the DF suppression to increase of number of particles in `co-rotating state', in which more field particles rotate with the GC in prograde direction than initial isotropic state. \citet{i:09}, however, revealed that the increase of co-rotating particles is marginal and anisotropy induced by the GC is very weak. Besides, \citet{gm:08} suggested that fluctuation of the halo potential weakens the DF in the core. But, they supposed the case of a black hole and a stellar cored structure, the perturber being so heavy that their theory is not applicable to the case of a dwarf galaxy and GCs.

\begin{figure}
 \includegraphics[width=84mm]{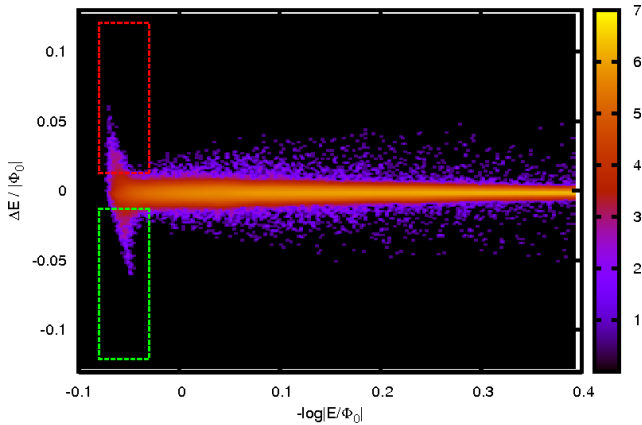}
 \caption{The same figure as Fig.\ref{fig3}, but the differential is taken between 7 and 8 Gyr: $\Delta E\equiv E(8Gyr)-E(7Gyr)$. Criteria for defining P-horn (N-horn) particles is $-\log|E/\Phi_0|<-3.0\times10^{-2}$ and $\Delta E/|\Phi_0|>1.27\times10^{-2}$ ($\Delta E/|\Phi_0|<-1.27\times10^{-2}$).}
 \label{T7-T8_contour}
\end{figure}
\begin{figure}
  \includegraphics[width=84mm]{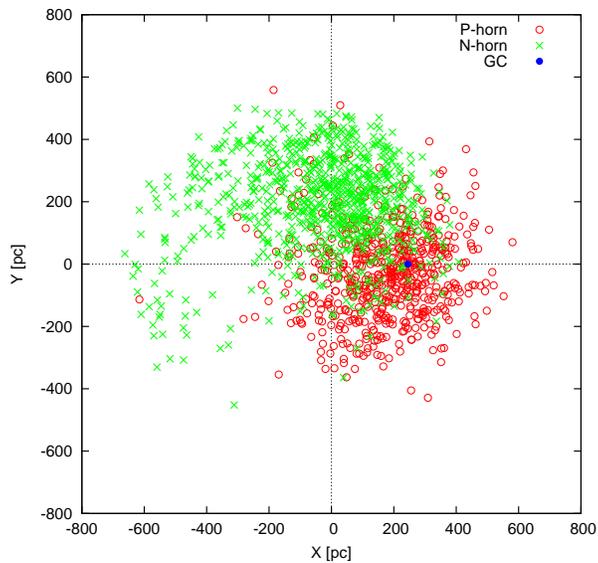}
  \caption{Positions of P-, N-horn particles and the GC at $t=7$ Gyr.}
  \label{T7-T8_positions}
\end{figure}
I execute the same analyses as in the previous subsection. In Fig.\ref{T7-T8_contour}, I redefine P- and N-horn particles by the same way. The numbers of the P- and N-horn particles are 613 and 834, respectively. Positions of these horn particles at t=7 Gyr are plotted in Fig.\ref{T7-T8_positions}. As well as the exerting DF phase, I confirm that most of the particles plotted have prograde rotations with the GC (399 of 613 P-horn and 761 of 834 N-horn particles).

\begin{figure}
  \includegraphics[width=84mm]{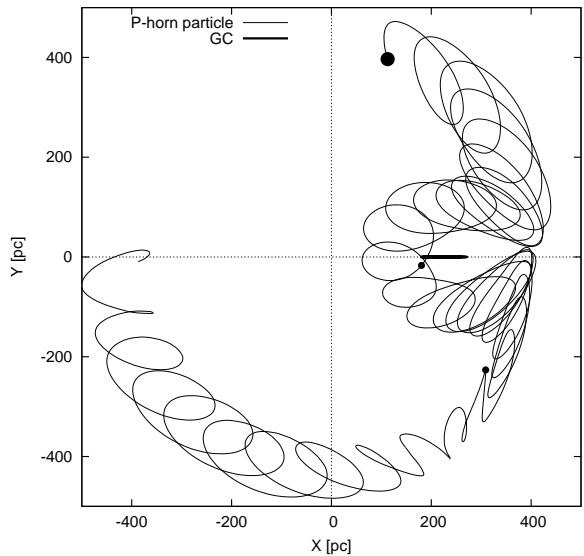}
  \includegraphics[width=84mm]{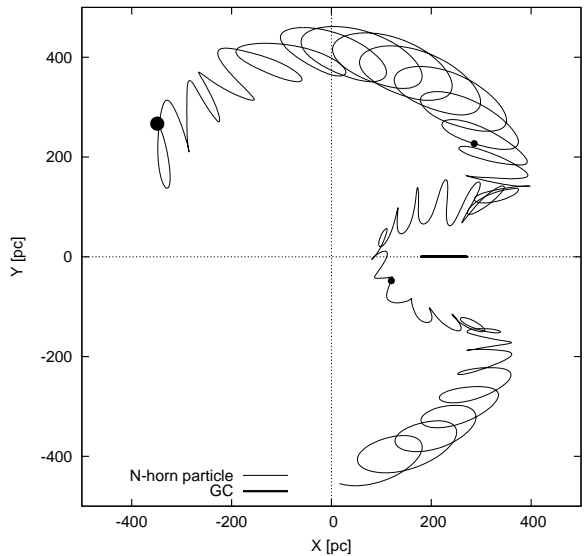}
  \caption{Examples of typical orbits of the P- and N-horn particle in the coordinate rotating with the GC. Although the particles are defined by energy increase in t=7-8 Gyr, the orbits plotted here are in t=6-9 Gyr. The large (small) dots indicate positions of the particles at t=6 Gyr (t=7 and t=8 Gyr).}
  \label{RP_T7}
\end{figure}
In Fig.\ref{RP_T7}, I illustrate typical orbits of P- and N-horn particle. This figure indicates very complex orbits. Both particles come close to the GC from the front, come inside the GC orbital radius, rotate around the GC, expand the orbits and finally go away behind the GC. It seems as if the particles dodge the GC. Although these particles are respectively defined as a P- and a N-horn particle, these orbits are identical. Regardless of the complexity, by considering absence of the donkey effect in the constant density region, the following explanation can be made. In constant densities, orbital period becomes independent of orbital radius. In coming close to the GC, the particles are orbiting slightly outside the core with slower angular velocities than the GC. By the gravity from the GC, the particles are pulled backward and lose energy. The particles sink toward the centre by the energy loss. However, by the absence of the donkey effect in the core region, the angular velocities of the particles are kept constant even in the inner radii. After getting behind the GC, the particles are pulled forward by the GC and gain energy. As a result, the orbits of the particles expand again and are exiled from the core. This orbit is also not in two-body interaction but in many-body interaction. In Appendix A, I will discuss about this orbit in restricted three-body problem. 

\begin{figure}
  \includegraphics[width=84mm]{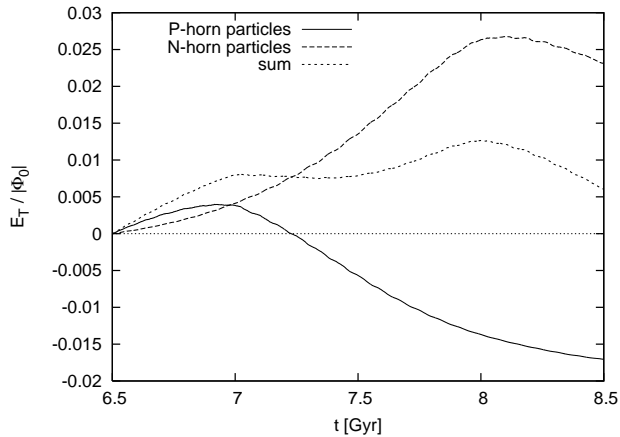}
  \caption{Cumulative energy transfers to the GC from the P- and N-horn particles in the suppressed DF phase ($t_{ini}=6.5$ Gyr in this plot).}
  \label{Et_horns_T7-T8}
\end{figure}
In Fig.\ref{Et_horns_T7-T8}, I show cumulative energy transfers into the GC. In consistency with the complexity of the orbits, the mechanical works on the GC also show very intricate features. The transfers are no longer monotonic, unlike in the exerting DF phase (Fig.\ref{fig5}). The N-horn particles continue to inject energy until $t\sim8$ Gyr, but absorb energy from the GC afterwards. Meanwhile, the P-horn particles inject a little energy at first, but largely absorb later. Thus, the total transfer from all horn particles indicates an intricate waving behavior. Although these particles are defined by $E(8Gyr)-E(7Gyr)$, the particles largely inject and absorb energy in $t\hspace{0.3em}\raisebox{0.4ex}{$<$}\hspace{-0.75em}\raisebox{-.7ex}{$\sim$}\hspace{0.3em}7$ Gyr and $t\hspace{0.3em}\raisebox{0.4ex}{$>$}\hspace{-0.75em}\raisebox{-.7ex}{$\sim$}\hspace{0.3em}8$ Gyr. These behaviors implies not transient but long-term interaction between the particles and the GC.

\begin{table}
  \begin{center}
    \caption{The numbers of particles defined as horn particles in each time-intervals. The top and bottom tables are 1Gyr-interval and 0.5Gyr-interval, respectively. The leftmost column indicates time-period in which I calculate $\Delta E$ to define the horn particles. The bottommost row is the number of particles for which all particles listed above are gathered. Because some particles overlap, the number of particle gathered is not identical to simple sum of rightmost columns.}
    \begin{tabular}{@{}lccr@{}}
      \hline\hline
      period     & P-horn & N-horn & total \\
      \hline
      5 - 6 Gyr  & 590 & 604 & 1194 \\
      6 - 7 Gyr  & 772 & 654 & 1426 \\
      7 - 8 Gyr  & 613 & 834 & 1447 \\
      8 - 9 Gyr  & 861 & 710 & 1571 \\
      9 - 10 Gyr & 929 & 884 & 1813 \\
      \hline
      gathered&  & & 4787 \\
      \hline
    \end{tabular}
    \begin{tabular}{@{}lccr@{}}
      \hline\hline
      period     & P-horn & N-horn & total \\
      \hline
      5 - 5.5 Gyr  &  195 &  345 &  540 \\
      5.5 - 6 Gyr  &  292 &  251 &  543 \\
      6 - 6.5 Gyr  &  253 &  337 &  590 \\
      6.5 - 7 Gyr  &  460 &  227 &  687 \\
      7 - 7.5 Gyr  &  365 &  462 &  827 \\
      7.5 - 8 Gyr  &  526 &  363 &  889 \\
      8 - 8.5 Gyr  &  977 &  352 & 1329 \\
      8.5 - 9 Gyr  &  696 &  658 & 1354 \\
      9 - 9.5 Gyr  & 1438 &  895 & 2333 \\
      9.5 - 10 Gyr & 1604 & 1082 & 2686 \\
      \hline
      gathered     &      &      & 6776 \\
      \hline
    \end{tabular}
  \label{list_1Gyr}
  \end{center}
\end{table}
\begin{figure}
  \includegraphics[width=84mm]{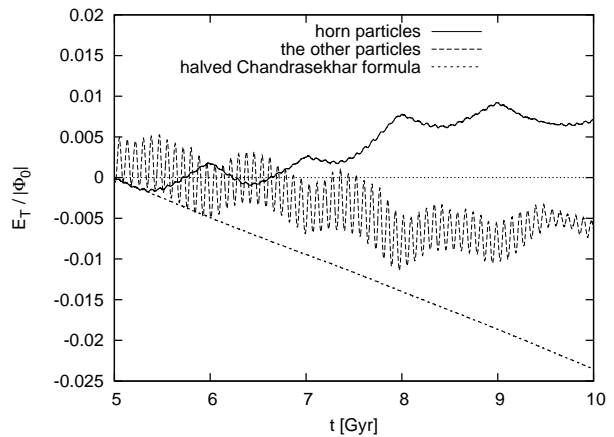}
  \caption{Cumulative energy transfers to the GC from the \textit{gathered} horn and the other particles in the suppressed DF phase by setting the time-interval to calculate $\Delta E$ to be 1 Gyr. For comparison, an analytical estimation by the Chandrasekhar formula is also plotted, but it is reduced by half from the naive estimation by Eq.\ref{chandra}.}
  \label{ET_1Gyr-interval}
\end{figure}

From the discussion above, in the suppressed DF phase, it is clear that the energy transfers from some particles have long duration. In this case, I couldn't extract all many-body interacting particles successfully. In order to overcome this difficulty, I screen particles in t=5-6, 6-7, 7-8, 8-9, 9-10 Gyr by calculating $\Delta E$.
If a particle is defined as a horn particle even only once in the five time-periods, I regard the particle as a many-body interacting particle in all time-periods. I gather all many-body interacting particles in this five time-periods. The numbers of particles listed up are shown in the top of Table \ref{list_1Gyr}. While operating the same simulation starting from t=5 Gyr, I compute energy transfers into the GC from the \textit{gathered} horn particles and the others.
 The result is shown in Fig.\ref{ET_1Gyr-interval}. In this figure, the cumulative energy transfer from the many-body interacting particles gradually increases. On the other hand, the other particles absorb energy from the GC. These cumulative transfers, however, indicate periodic flapping. This periodicity appears just in 1 Gyr period, implying an artificial effect caused by setting the time-interval to calculate $\Delta E$ to be 1 Gyr. 

\begin{figure}
  \includegraphics[width=84mm]{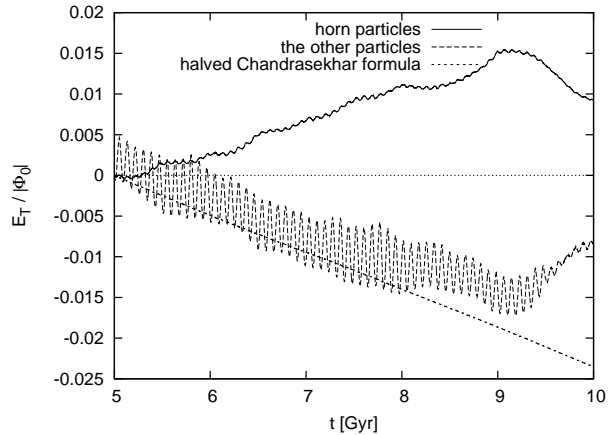}
  \caption{The same plot as Fig.\ref{ET_1Gyr-interval}, but using 0.5Gyr-interval to extract the gathered horn particles.}
  \label{ET_0.5Gyr-interval}
\end{figure}
Therefore, I follow the same procedure to calculate $\Delta E$ with 0.5Gyr-interval. The numbers of particles listed up are shown in the bottom of Table \ref{list_1Gyr}. I show cumulative energy transfers in Fig.\ref{ET_0.5Gyr-interval} with an analytical estimation by Chandrasekhar formula for the sake of comparison. However, the analytical estimation is halved, according to the result of \S 3.1. The periodic flapping in Fig.\ref{ET_1Gyr-interval} disappears. Moreover, the energy injection from the horn particles are monotonic for $t\hspace{0.3em}\raisebox{0.4ex}{$<$}\hspace{-0.75em}\raisebox{-.7ex}{$\sim$}\hspace{0.3em}9$ Gyr. The energy absorption by the other particles is also monotonic. Interestingly, the behavior of the energy transfer from the horn particles is largely different from that in the exerting DF phase; the horn, many-body interacting, particles continue to inject energy into the GC in the suppressed DF phase. On the other hand, the other particles absorb energy from the GC as well as in the exerting DF phase. The halved analytic formula provides excellent fit to the absorption for $t\hspace{0.3em}\raisebox{0.4ex}{$<$}\hspace{-0.75em}\raisebox{-.7ex}{$\sim$}\hspace{0.3em}9$ Gyr though there is a slight deviation at $t\sim 8$ Gyr. These influences from horn and the other particles are canceled out and the GC orbit can remain on a constant orbital radius as if the DF on the GC vanished.



\section{discussion and summary}

By the N-body simulation in this paper, I conclude as follows. In the exerting DF phase, the Chandrasekhar formula seems good at estimating the actual DF. Nevertheless, quantitatively, a very small number of particles accounts for a non-negligible fraction of the actual DF although they occupy less than 0.1 per cent of the halo mass. These particles are in many-body interaction, which have quite different orbits from two-body interaction. The presence of such many-body interacting particles disagrees with the concept of \citet{c:43}. However, in his formula, uncertainty of determining $\ln\Lambda$ would conceal the contribution from these particles.

In the suppressed DF phase, the many-body interacting particles alter direction of energy transfer and feed energy into the perturber. On the other hand, the other particles exert DF, the orthodox DF, ceaselessly. In this phase, as a result, these energy transfers cancel out with each other. The orbit of the perturber is thus stabilized. I suggest that this could be a possible mechanism of the suppressed DF in a constant density structure. 

Although my results indicate a failure of the Chandrasekhar formula, this is not surprising. The derivation of the formula has assumed an unrealistic circumstance: uniform density, isotropic velocity field and non-periodic straight orbit. Moreover, a perturber in a real galaxy has to rotate around center of the galaxy. Therefore, from the very beginning it is unreasonable to apply the Chandrasekhar formula to all kinds of density distributions.

\citet{tw:84} and \citet{w:89} have analytically suggested energy feedback on a perturber by density response of field. Especially, figures in \citet{w:89} are very intelligible schematically. He indicated shapes of density wakes. In his results, outside core region, the density wake is asymmetric and exerts drag force on the perturber. On the other hand, inside the core, the wake become symmetric and the drag force disappears. The symmetric shape is similar to the distribution of the horn particles in my Fig.\ref{T7-T8_positions} although the horn particles in my simulation is selected by energy response, not density. However, in my simulation, the symmetric density wake indicated in \citet{w:89} can not be observed in the suppressed DF phase. Besides, the N-body simulations of \citet{rgm:06}, almost the same simulations as mine, also don't indicate the symmetric density wake, but asymmetric wake (see Fig.4 of \citet{rgm:06}) even inside the core region. \citet{gmr:10} also showed similar asymmetric desity wake in their Fig.5. This discrepancy may or may not be caused by low resolution in our simulations. Also, \citet{bpp:08} has suggested that the field particles on circular orbits lose or gain energy according to the relative phase to the orbit of the perturber while the perturber plays the role of a catalyst to modify the distribution function of the field particles and induces anisotropy. 

The result of this paper indicates that not only two-body interactions but also many-body interactions are responsible for the actual DF. The many-body interacting particles, horn particles, are very small in number and co-rotating with the perturber. Moreover, the resistance to the DF by this corrective effect is also feasible in shallow cusp density profiles although the shallow cusp may evolve into a constant density (see, Appendix A). This implies sensitivity of the actual DF to velocity field of the halo models. I employed in this paper the isotropic assumption, $\beta=0.0$ in Eq.\ref{jeans}, as the most fundamental model. However, many numerical simulations have suggested dependence of $\beta$ on radius in dwarf and other galaxies \citep{a:82,mwc:08,bt:08}. These results imply that actual DF against a \textit{live} halo would depend sensitively on the distribution functions of the haloes. In addition, there are some generalised applications of the Chandrasekhar formula; against aspherical distributions \citep{b:77,pcv:92}, variable $\ln\Lambda$ \citep{hfm:03}, DF on spherical bodies \citep{ef:07}, in a gaseous medium\citep[e.g.][]{o:99,kks:08}, etc. The results in this paper alert to thoughtless usages of the Chandrasekhar formula.

\section*{Acknowledgments}

 The author is financially supported by Research Fellowships of the Japan Society for the Promotion of Science (JSPS) for Young Scientists. The numerical simulations reported here were carried out on MUV (Mitaka Underground Vineyard) GRAPE systems kindly made available by CfCA (Center for Computational Astrophysics) at National Astronomical Observatory of Japan. The numerical code was based on a software distributed on the website of Joshua E. Barnes (http://ifa.hawaii.edu/\~{}barnes/software.html). I thank Masafumi Noguchi for his fascinating and helpful discussion, Masaki Iwasawa and Yusuke Tsukamoto for innovative advices. It is pleasure to thank Martin D. Weinberg for his stimulating comments, which drastically improved the description given in this paper and my understanding of galactic dynamics. I thank the anonymous reviewer for important suggestions that helped improve the paper.


\begin{thebibliography}{}

\bibitem[\protect\citeauthoryear{Angus \& Diaferio}{Angus \&
  Diaferio}{2009}]{ad:09}
Angus G.~W.,  Diaferio A.,  2009, MNRAS, 396, 887

\bibitem[\protect\citeauthoryear{Barnes}{Barnes}{1990}]{b:90}
Barnes J.,  1990, J. Comput. Phys., 87, 161

\bibitem[\protect\citeauthoryear{Barnes \& Hut}{Barnes \& Hut}{1986}]{bh:86}
Barnes J.,  Hut P.,  1986, Nat, 324, 446

\bibitem[\protect\citeauthoryear{Bekki}{Bekki}{2010}]{b:09}
Bekki K.,  2010, MNRAS, 401, 2753

\bibitem[\protect\citeauthoryear{Binney}{Binney}{1977}]{b:77}
Binney J.,  1977, MNRAS, 181, 735

\bibitem[\protect\citeauthoryear{Binny \& Tremaine}{Binny \&
  Tremaine}{2008}]{bt:08}
Binny J.,  Tremaine S.,  2008, Galactic Dynamics Second Edition.
Princeton Univ. Press, Princeton

\bibitem[\protect\citeauthoryear{Boily, Padmanabhan \& Paiement}{Boily
  et~al.}{2008}]{bpp:08}
Boily C.~M.,  Padmanabhan T.,    Paiement A.,  2008, MNRAS, 383, 1619

\bibitem[\protect\citeauthoryear{Buonanno, Corsi, Castellani, Marconi, Pecci \&
  Zinn}{Buonanno et~al.}{1999}]{bcc:99}
Buonanno R.,  Corsi C.~E.,  Castellani M.,  Marconi G.,  Pecci F.~F.,    Zinn
  R.,  1999, AJ, 118, 1671

\bibitem[\protect\citeauthoryear{Buonanno, Corsi, Zinn, Pecci, Hardy \&
  Suntzeff}{Buonanno et~al.}{1998}]{bcz:98}
Buonanno R.,  Corsi C.~E.,  Zinn R.,  Pecci F.~F.,  Hardy E.,    Suntzeff
  N.~B.,  1998, ApJ, 501, 33

\bibitem[\protect\citeauthoryear{Burkert}{Burkert}{1995}]{b:95}
Burkert A.,  1995, ApJ, 447, 25

\bibitem[\protect\citeauthoryear{Capuzzo-Dolcetta \& Miocchi}{Capuzzo-Dolcetta
  \& Miocchi}{2008}]{cm:08}
Capuzzo-Dolcetta R.,  Miocchi P.,  2008, ApJ, 681, 1136

\bibitem[\protect\citeauthoryear{Chandrasekhar}{Chandrasekhar}{1943}]{c:43}
Chandrasekhar S.,  1943, ApJ, 97, 255

\bibitem[\protect\citeauthoryear{Cowsik, Wagoner, Berti \& Sircar}{Cowsik
  et~al.}{2009}]{cwb:09}
Cowsik R.,  Wagoner K.,  Berti E.,    Sircar A.,  2009, ApJ, 699, 1389

\bibitem[\protect\citeauthoryear{{de Blok}}{{de Blok}}{2010}]{d:10}
{de Blok} W.~J.~G.,  2010, Adv. Astron., 2010, 789293

\bibitem[\protect\citeauthoryear{de Blok, McGaugh, Bosma \& Rubin}{de~Blok
  et~al.}{2001}]{bmb:01}
de Blok W. J.~G.,  McGaugh S.~S.,  Bosma A.,    Rubin V.~C.,  2001, ApJ, 552,
  23

\bibitem[\protect\citeauthoryear{Durrell, Harris, Geisler \& Pudritz}{Durrell
  et~al.}{1996}]{dhg:96}
Durrell P.~R.,  Harris W.~E.,  Geisler D.,    Pudritz R.~E.,  1996, AJ, 112,
  972

\bibitem[\protect\citeauthoryear{Esquivel \& Fuchs}{Esquivel \&
  Fuchs}{2007}]{ef:07}
Esquivel O.,  Fuchs B.,  2007, MNRAS, 378, 1191

\bibitem[\protect\citeauthoryear{Fujii, Funato \& Makino}{Fujii
  et~al.}{2006}]{ffm:06}
Fujii M.,  Funato Y.,    Makino J.,  2006, PASJ, 58, 743

\bibitem[\protect\citeauthoryear{Fujii, Iwasawa, Funato \& Makino}{Fujii
  et~al.}{2009}]{fif:09}
Fujii M.,  Iwasawa M.,  Funato Y.,    Makino J.,  2009, ApJ, 695, 1421

\bibitem[\protect\citeauthoryear{Gilmore, Wilkinson, Wyse, Kleyna, Koch, Evans
  \& Grebel}{Gilmore et~al.}{2007}]{gww:07}
Gilmore G.,  Wilkinson M.~I.,  Wyse R. F.~G.,  Kleyna J.~T.,  Koch A.,  Evans
  N.~W.,    Grebel E.~K.,  2007, ApJ, 663, 948

\bibitem[\protect\citeauthoryear{{Goerdt}, {Moore}, {Read} \&
  {Stadel}}{{Goerdt} et~al.}{2010}]{gmr:10}
{Goerdt} T.,  {Moore} B.,  {Read} J.~I.,    {Stadel} J.,  2010, ApJ, 725, 1707

\bibitem[\protect\citeauthoryear{Goerdt, Moore, Read, Stadel \& Zemp}{Goerdt
  et~al.}{2006}]{gmr:06}
Goerdt T.,  Moore B.,  Read J.~I.,  Stadel J.,    Zemp M.,  2006, MNRAS, 368,
  1073

\bibitem[\protect\citeauthoryear{{Governato} et~al.,}{{Governato}
  et~al.}{2010}]{g:10}
{Governato} F.,  et~al., 2010, Nat, 463, 203

\bibitem[\protect\citeauthoryear{Greco et~al.,}{Greco  et~al.}{2007}]{g:07}
Greco C.,  et~al., 2007, ApJ, 670, 332

\bibitem[\protect\citeauthoryear{Gualandris \& Merritt}{Gualandris \&
  Merritt}{2008}]{gm:08}
Gualandris A.,  Merritt D.,  2008, ApJ, 678, 780

\bibitem[\protect\citeauthoryear{Hashimoto, Funato \& Makino}{Hashimoto
  et~al.}{2003}]{hfm:03}
Hashimoto Y.,  Funato Y.,    Makino J.,  2003, ApJ, 582, 196

\bibitem[\protect\citeauthoryear{Hernandez \& Gilmore}{Hernandez \&
  Gilmore}{1998}]{hg:98}
Hernandez X.,  Gilmore G.,  1998, MNRAS, 297, 517

\bibitem[\protect\citeauthoryear{Hernquist}{Hernquist}{1993}]{h:93}
Hernquist L.,  1993, ApJ, 86, 389

\bibitem[\protect\citeauthoryear{Inoue}{Inoue}{2009}]{i:09}
Inoue S.,  2009, MNRAS, 397, 709

\bibitem[\protect\citeauthoryear{Jiang \& Binney}{Jiang \&
  Binney}{2000}]{jb:00}
Jiang I.-G.,  Binney J.,  2000, MNRAS, 314, 468

\bibitem[\protect\citeauthoryear{Kim, Kim \& S\'anchez-Salcedo}{Kim
  et~al.}{2008}]{kks:08}
Kim H.,  Kim W.-T.,    S\'anchez-Salcedo F.~J.,  2008, ApJ, 679, 33

\bibitem[\protect\citeauthoryear{{Li} \& {Chen}}{{Li} \& {Chen}}{2009}]{lc:09}
{Li} N.,  {Chen} D.,  2009, Research in Astronomy and Astrophysics, 9, 1173

\bibitem[\protect\citeauthoryear{{Liu}, {Fu}, {Deng} \& {Huang}}{{Liu}
  et~al.}{2005}]{lfd:05}
{Liu} W.,  {Fu} Y.,  {Deng} Z.,    {Huang} J.,  2005, PASJ, 57, 541

\bibitem[\protect\citeauthoryear{Lotz, Miller \& Ferguson}{Lotz
  et~al.}{2004}]{lmf:04}
Lotz J.~M.,  Miller B.~W.,    Ferguson H.~C.,  2004, ApJ, 613, 262

\bibitem[\protect\citeauthoryear{Lynden-Bell \& Kalnajs}{Lynden-Bell \&
  Kalnajs}{1972}]{lk:72}
Lynden-Bell D.,  Kalnajs A.~J.,  1972, MNRAS, 157, 1

\bibitem[\protect\citeauthoryear{Mackey \& Gilmore}{Mackey \&
  Gilmore}{2003}]{mg:03}
Mackey A.~D.,  Gilmore G.~F.,  2003, MNRAS, 340, 175

\bibitem[\protect\citeauthoryear{Makino}{Makino}{1991}]{m:91}
Makino J.,  1991, PASJ, 43, 621

\bibitem[\protect\citeauthoryear{Mashchenko, Couchman \& Wadsley}{Mashchenko
  et~al.}{2006}]{mcw:06}
Mashchenko S.,  Couchman H. M.~P.,    Wadsley J.,  2006, Nat, 442, 539

\bibitem[\protect\citeauthoryear{Mashchenko, Wadsley \& Couchman}{Mashchenko
  et~al.}{2008}]{mwc:08}
Mashchenko S.,  Wadsley J.,    Couchman H. M.~P.,  2008, Sci, 319, 174

\bibitem[\protect\citeauthoryear{Miller, Lotz, Ferguson, Stiavelli \&
  Whitmore}{Miller et~al.}{1998}]{mlf:98}
Miller B.~W.,  Lotz J.~M.,  Ferguson H.~C.,  Stiavelli M.,    Whitmore B.~C.,
  1998, ApJ, 508, 133

\bibitem[\protect\citeauthoryear{Miocchi, Capuzzo-Dolcetta, {Di Matteo} \&
  Vicari}{Miocchi et~al.}{2006}]{mcd:06}
Miocchi P.,  Capuzzo-Dolcetta R.,  {Di Matteo} P.,    Vicari A.,  2006, ApJ,
  644, 940

\bibitem[\protect\citeauthoryear{Navarro, Frenk \& White}{Navarro
  et~al.}{1997}]{nfw:97}
Navarro J.~F.,  Frenk C.~S.,    White S. D.~M.,  1997, ApJ, 490, 493

\bibitem[\protect\citeauthoryear{Oh, Lin \& Richer}{Oh et~al.}{2000}]{olr:00}
Oh K.~S.,  Lin D. N.~C.,    Richer H.~B.,  2000, ApJ, 531, 727

\bibitem[\protect\citeauthoryear{{Oh}, {Brook}, {Governato}, {Brinks}, {Mayer},
  {de Blok}, {Brooks} \& {Walter}}{{Oh} et~al.}{2010}]{obg:10}
{Oh} S.,  {Brook} C.,  {Governato} F.,  {Brinks} E.,  {Mayer} L.,  {de Blok}
  W.~J.~G.,  {Brooks} A.,    {Walter} F.,  2010, ArXiv 1011.2777

\bibitem[\protect\citeauthoryear{Ostriker}{Ostriker}{1999}]{o:99}
Ostriker E.~C.,  1999, ApJ, 513, 252

\bibitem[\protect\citeauthoryear{{Pasetto}, {Grebel}, {Berczik}, {Spurzem} \&
  {Dehnen}}{{Pasetto} et~al.}{2010}]{pgb:10}
{Pasetto} S.,  {Grebel} E.~K.,  {Berczik} P.,  {Spurzem} R.,    {Dehnen} W.,
  2010, {A\&A}, 514, 47

\bibitem[\protect\citeauthoryear{Pesce, Capuzzo-Dolcetta \& Vietri}{Pesce
  et~al.}{1992}]{pcv:92}
Pesce E.,  Capuzzo-Dolcetta R.,    Vietri M.,  1992, MNRAS, 254, 466

\bibitem[\protect\citeauthoryear{Read, Goerdt, Moore, pontzen \& Stadal}{Read
  et~al.}{2006}]{rgm:06}
Read J.~I.,  Goerdt T.,  Moore B.,  pontzen A.~P.,    Stadal J.,  2006, MNRAS,
  373, 1451

\bibitem[\protect\citeauthoryear{Salucci \& Burkert}{Salucci \&
  Burkert}{2000}]{sb:00}
Salucci P.,  Burkert A.,  2000, ApJ, 537, 9

\bibitem[\protect\citeauthoryear{S\'anchez-Salcedo, Reyes-Iturbide \&
  Hernandez}{S\'anchez-Salcedo et~al.}{2006}]{srh:06}
S\'anchez-Salcedo F.~J.,  Reyes-Iturbide J.,    Hernandez X.,  2006, MNRAS,
  370, 1829

\bibitem[\protect\citeauthoryear{Springel et~al.,}{Springel
  et~al.}{2008}]{swv:08}
Springel V.,  et~al., 2008, MNRAS, 391, 1685

\bibitem[\protect\citeauthoryear{Springel \& White}{Springel \&
  White}{1999}]{sw:99}
Springel V.,  White D.~M.,  1999, MNRAS, 307, 162

\bibitem[\protect\citeauthoryear{Strader, Brodie, Forbes, Beasley \&
  Huchra}{Strader et~al.}{2003}]{sbf:03}
Strader J.,  Brodie J.~P.,  Forbes D.~A.,  Beasley M.~A.,    Huchra J.~P.,
  2003, AJ, 125, 1291

\bibitem[\protect\citeauthoryear{Strigari, Bullock, Kaplinghat, Simon, Geha,
  Willman \& Walker}{Strigari et~al.}{2008}]{sbk:08}
Strigari L.~E.,  Bullock J.~S.,  Kaplinghat M.,  Simon J.~D.,  Geha M.,
  Willman B.,    Walker M.~G.,  2008, Nat, 454, 1096

\bibitem[\protect\citeauthoryear{Tremaine}{Tremaine}{1976}]{t:76}
Tremaine S.,  1976, ApJ, 203, 345

\bibitem[\protect\citeauthoryear{Tremaine \& Weinberg}{Tremaine \&
  Weinberg}{1984}]{tw:84}
Tremaine S.,  Weinberg M.~D.,  1984, MNRAS, 209, 729

\bibitem[\protect\citeauthoryear{van Albada}{van Albada}{1982}]{a:82}
van Albada T.~S.,  1982, MNRAS, 201, 939

\bibitem[\protect\citeauthoryear{Vesperini}{Vesperini}{2000}]{v:00}
Vesperini E.,  2000, MNRAS, 318, 841

\bibitem[\protect\citeauthoryear{Vesperini}{Vesperini}{2001}]{v:01}
Vesperini E.,  2001, MNRAS, 322, 247

\bibitem[\protect\citeauthoryear{{Weinberg}}{{Weinberg}}{1989}]{w:89}
{Weinberg} M.~D.,  1989, MNRAS, 239, 549

\end{thebibliography}

\appendix

\section[]{circular restricted three-body problem in the core region}
I discuss here why and how the orbits indicated in Fig.\ref{RP_T7} appear in the suppressed DF phase. Since the orbit of the perturber is nearly circular in my simulation and density field is stable, I expect that circular restricted three-body problem would be a valid method to investigate the orbits in the core region. Here, I assume that the GC is orbiting at a separation of $R_0=300pc$ from the galactic center in a circular orbit. Let the enclosed mass of the halo, $M_0\equiv 4\pi\int^{R_0}_0\rho(r)r^2dr$. I set the coordinate center to the center of mass of the system. Then, the galactic center and the GC are located at $\vec{x}_M=[-R_0m_{gc}/(M_0+m_{gc}),0,0]$ and $\vec{x}_{gc}=[R_0M_0/(M_0+m_{gc}),0,0]$, respectively. Then, I calculate effective potential at $\vec{x}=[x,y,0]$ in the coordinate rotating with the GC,
\begin{equation}
  \Phi_{eff}(\vec{x})=\Phi(R_M)-G\Bigl[\frac{m_{gc}}{R_{gc}}+\frac{M_0+m_{gc}}{2R_0^3}(x^2+y^2)\Bigr],
  \label{jacobi}
\end{equation}
where $R_M\equiv|\vec{x}-\vec{x}_M|$, $R_{gc}\equiv|\vec{x}-\vec{x}_{gc}|$. The first and second term are gravitational potential of the halo density distribution and the GC, respectively. The third term is centrifugal potential. 
 
\begin{figure}
  \includegraphics[width=100mm]{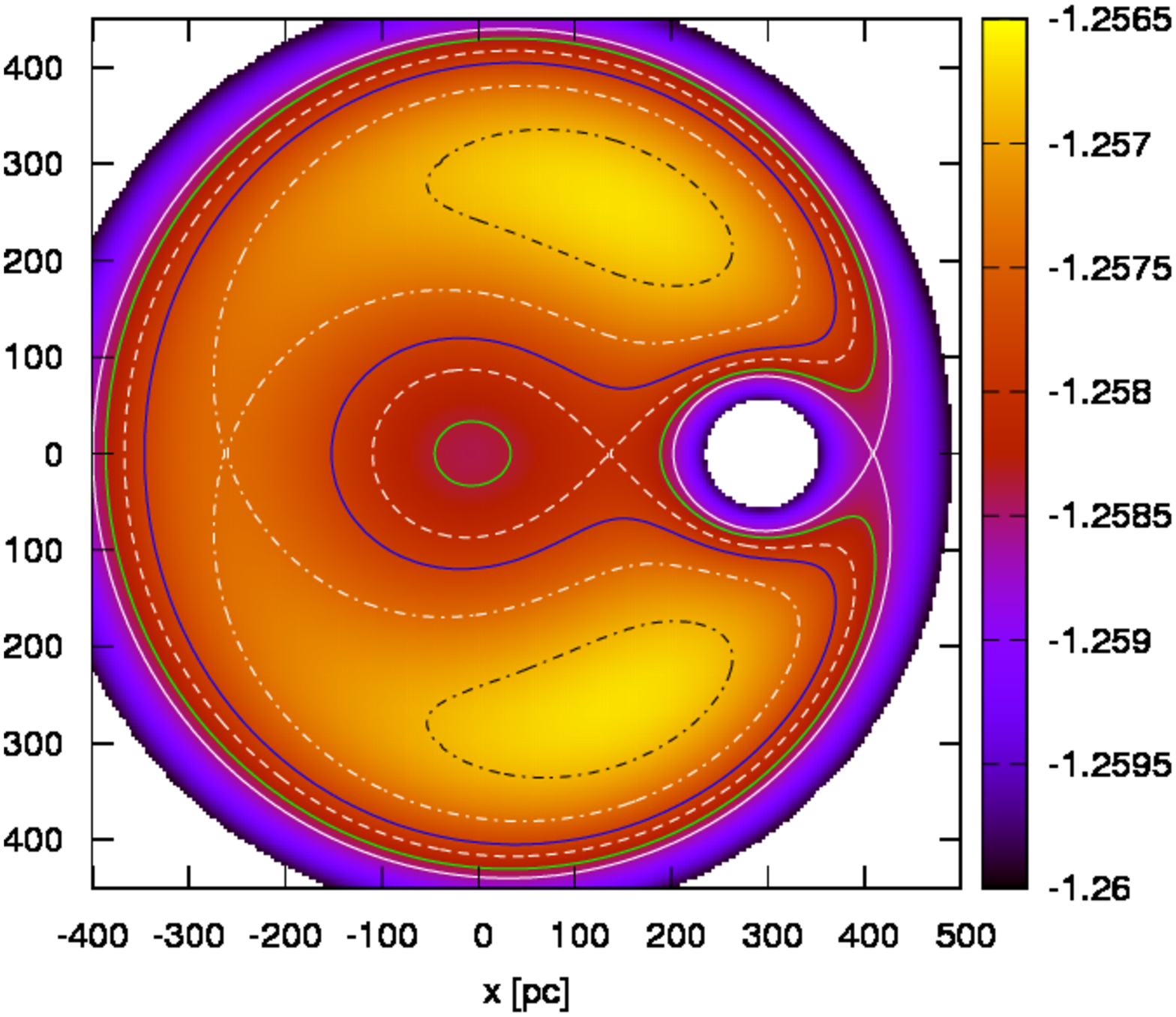}
  \caption{Contour plot of effective potential in the case where the perturber is rotating inside the core region with $r_0=1kpc$. Colour-bar labels the energy, $\Phi_{eff}/|\Phi_0|$. The coordinate center is the center of mass of the system. The dot-dashed, solid and dashed white lines indicate equipotential surfaces of Lagrangian points L1, L2 and L3, respectively. The green line indicates an isosurface on which the orbits indicated in Fig.\ref{RP_T7} are possible.}
  \label{Mas}
\end{figure}

In Fig.\ref{Mas}, I show a contour plot of the effective potential in the case of that the core radius is set to $r_0=1kpc$, which corresponds to the suppressed DF phase in my simulation. I also plot equipotential surfaces of Lagrangian points, L1, L2 and L3. The green line indicates a equipotential surface between the effective potentials at L2 and L3: $\Phi_{eff,L2}$ and $\Phi_{eff,L3}$. This green line runs through around the GC just like the horn-particle orbits in Fig.\ref{RP_T7}. If field particles have a Jacobi integral higher than $\Phi_{eff,L2}$ and lower than $\Phi_{eff,L3}$, some of these can orbit along this green line. I consider the particles shown in Fig.\ref{RP_T7} to be examples of this kind of particle.

Special attention should be paid to the fact that the effective potential at L2 (the solid white line) is lower than that at L3 (the dashed white line). This height relation between $\Phi_{eff,L2}$ and $\Phi_{eff,L3}$ permits appearance of orbits like Fig.\ref{RP_T7}. On the other hand, in cases of cuspy density distribution, $\Phi_{eff,L2}$ is generally higher than $\Phi_{eff,L3}$ (see below).

In Fig.\ref{Tadpole}, I show a contour plot of the effective potential in the case where the core radius is set to $r_0=50pc$, which corresponds to the case of cuspy density distribution: the exerting DF phase. In this case, $\Phi_{eff,L2}$ is higher than $\Phi_{eff,L3}$. Then, the green lines in this case can \textit{not} run across between potential minima, \textit{i.e.}, no orbit can run through between the GC and the galactic center since such a route is completely prohibited. The green lines in Fig.\ref{Tadpole} correspond to the horn-particle orbits in Fig.\ref{RP_T2}.

\label{lastpage}\begin{figure}
  \includegraphics[width=100mm]{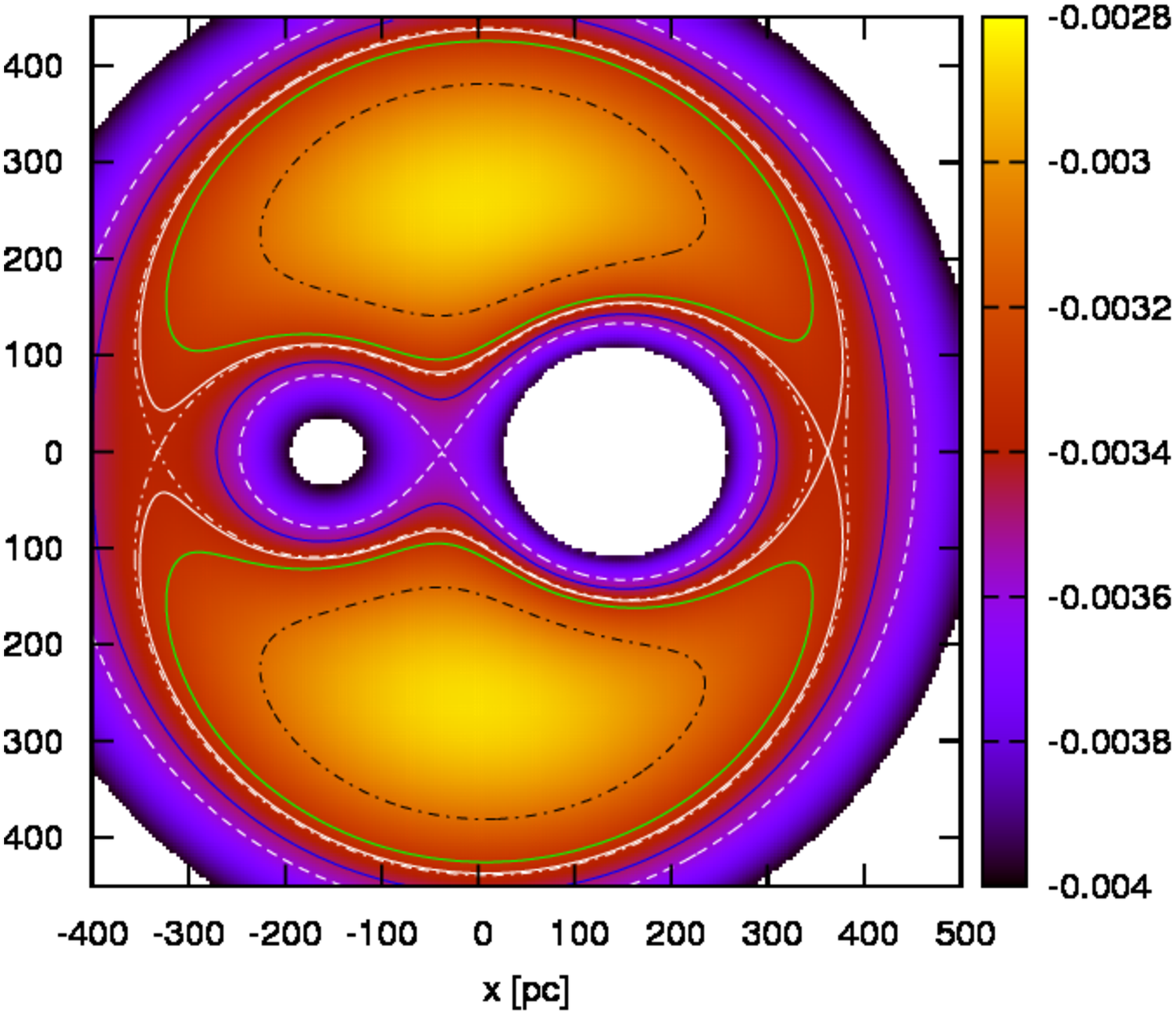}
  \caption{Contour plot of effective potential in the case of cusy density distribution with $r_0=50pc$. The dot-dashed, solid and dashed white lines indicate equipotential surfaces of Lagrangian points L1, L2 and L3, respectively. The green line indicates an isosurface on which the orbits indicated in Fig.\ref{RP_T2} are possible.}
  \label{Tadpole}
\end{figure}

In the case of the cored density, if the GC is orbiting in even inner region of the core, much more particles can become horn-particles shown as the green line in Fig.\ref{Mas}. Therefore, force against DF becomes strong toward inner region, at a certain radius, the force balances with the DF on the GC. This means that the DF suppression is not attributed anisotropic velocity state induced by the GC rotation: co-rotating state suggested by \citet{rgm:06}. \citet{i:09} has shown that such a co-rotating anisotropic velocity state is very marginal. If the appearance of horn-particles is the true mechanism of DF suppresion, even though the velocity distribution is isotropic, core-stalling of DF is possible. Although the horn-particles I suggest are co-rotating particlles in a sense, but \textit{there is no need to induce the anisotropy} in the velocity field. This is consistent with the marginally anisotropic velocity distribution found by \citet{i:09}.

Next, I address the cases of shallow cusp density distibutions by means of the circular restricted three-body problem. Some studies have concluded that actual dark matter profile of dwarf galaxies would have a shallow inner cusp with slopes between 0.3-0.6 \citep{g:10,obg:10}. In order to investigate the cases of shallow cusps, I use another density profile,
\begin{equation}
  \rho(r)=\frac{\rho_0}{(r/r_0)^\gamma\bigl[1+(r/r_0)^\alpha\bigr]^{(\beta-\gamma)/\alpha}},
\end{equation}
where $\rho_0 = 0.1M_\odot/pc^3$ and $r_0=1.0kpc$. I set the parameters, $\alpha=1.5$, $\beta=3.0$ (outer slope) and $\gamma=0.3$ or $0.6$ (inner slope). If $\alpha=1.5$, $\beta=3.0$ and $\gamma=0.0$, this profile becomes very similar to Burkert profile. In this profile, I assume that the GC with the mass of $m_{gc} = 2.0\times10^5M_\odot$ is orbiting at a separation of $R_0=300pc$ from the galactic center in a circular orbit. 

\begin{figure}
 \includegraphics[width=84mm]{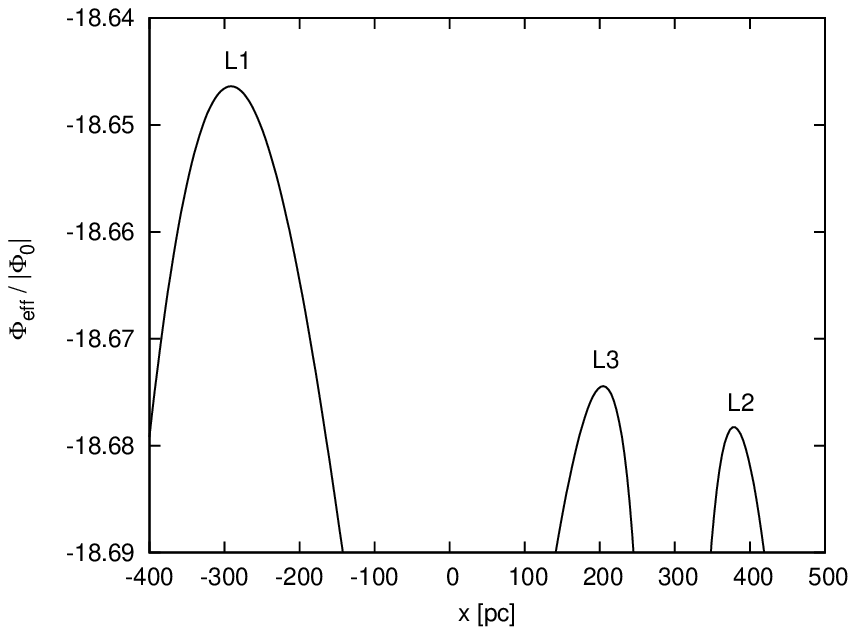}
 \caption{Sectional effective potential surface on x-axis (on the line of the GC and halo centre) in the case of $\gamma=0.3$.}
 \label{A1}
\end{figure}

As I discussed above, some field particles can be in the special orbit (Fig.\ref{RP_T7}) if the effective potential on L3 point is higher than that on L2. This fact implies that all I have to do is to calculate the effective potential on L3 and L2. Then, I plot the effective potential along x-axis (the line of the GC and the halo centre). Fig.\ref{A1} shows the effective potential in the case of $\gamma=0.3$. In this case, the effective potential on L3 is higher than that on L2. I expect that some horn particles in the special orbit would be existent. Then, according to the main discussion of the paper, these horn particles could counter and weaken DF on the GC.

\begin{figure}
 \includegraphics[width=84mm]{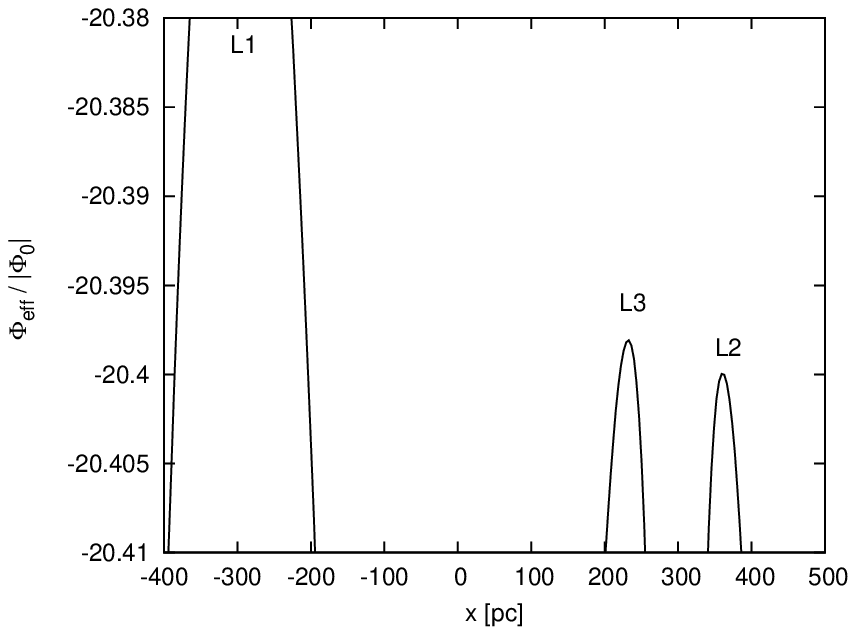}
 \caption{Sectional effective potential surface on x-axis in the case of $\gamma=0.6$.}
 \label{A2}
\end{figure}

I show the effective potential in the case of $\gamma=0.6$ in Fig.\ref{A2}. The position of the GC, the scale radius, $r_0$, and other parameters are the same. The effective potential on L3 is still higher than that on L2. However, the difference between these is very small in comparison with the previous case. Only a minority of particles can be in the special orbit if $\gamma=0.6$. In this case, the horn particles may be unable to balance against the DF on the GC.

From the discussion above, steeper inner slope of the density profile allows fewer particles to orbit in many-body interaction manner like Fig.\ref{RP_T7}. However, in this discussion, it is very difficult or impossible to investigate how many particles in this orbit are required to cancel out the DF on the GC. After all, detailed investigation needs to operate N-body simulations. However, as \citet{rgm:06} has concluded, N-body simulation suffers from the decline of the inner cusp. Then, I can not conclude what value of $\gamma$ is the criterion of the DF stalling in (nearly) cored density profiles. Since steeper (shallower) cusp allows fewer (more) horn particles, the number of particles required to observe this kind of orbits in N-body simulations would depend on the inner slope. In the cases of steeper cusps, to operate N-body simulation needs more particles.

\end{document}